# Oxygen vacancy induced room temperature metal-insulator transition in nickelates films and its potential application in photovoltaics

Le Wang[†], Sibashisa Dash[†], Lei Chang[†], Lu You[†], Yaqing Feng[‡], Xu He[‡], Kui-juan Jin[‡,#], Yang Zhou[†], Hock Guan Ong[§], Peng Ren[†], Shiwei Wang[†], Lang Chen[‖], and Junling Wang[†,*]

[†] School of Materials Science and Engineering, Nanyang Technological University, Singapore 639798, Singapore

[‡] Institute of Physics, Chinese Academy of Sciences, Beijing 100190, China

[#] Collaborative Innovation Center of Quantum Matter, Beijing, 100190, China

[§] Temasek Laboratories@NTU, Nanyang Technological University, Singapore 637553, Singapore

[‖] Department of Physics, South University of Science and Technology of China, Shen Zhen, 518055, China

**ABSTRACT:** Oxygen vacancy is intrinsically coupled with magnetic, electronic and transport properties of transition-metal oxide materials and directly determines their multifunctionality. Here, we demonstrate reversible control of oxygen content by post-annealing at temperature lower than 300℃ and realize the reversible metal-insulator transition in epitaxial NdNiO$_3$ films. Importantly, over six orders of magnitude in the resistance modulation and a large change in optical band gap are demonstrated at room temperature without destroying the parent framework and changing the p-type conductive mechanism. Further study revealed that oxygen vacancies stabilized the insulating phase at room temperature is universal for perovskite nickelates films. Acting as electron donors, oxygen vacancies not only stabilize the insulating phase at room temperature, but also induce a large magnetization of ~50 emu/cm$^3$ due to the formation of strongly correlated Ni$^{2+}$ $t_{2g}^6 e_g^2$ states. The band gap opening is an order of magnitude larger than that of the thermally driven metal-insulator transition and continuously tunable. Potential application of the newly found insulating phase in photovoltaics has been demonstrated in the nickelates-based heterojunctions. Our discovery opens up new possibilities for strongly correlated perovskite nickelates.

**INTRODUCTION**

Transition-metal oxides possess a wide spectrum of electronic and magnetic properties, including high-temperature superconductivity, multiferroicity and Mott metal-insulator transition (MIT).[1-3] The multiple nearly degenerate ground states render them extremely sensitive to perturbations, ideal for understanding various quantum physics of strongly correlated electrons. In particular, the rich phase diagrams of perovskite nickelates ($RNiO_3$, where R represents rare earth lanthanide elements) are of great importance for both materials physics and oxide electronics.[4-11] They usually show a thermally driven first order MIT at temperature $T_{MI}$ and a paramagnetic-antiferromagnetic transition at the Neel temperature ($T_N$). The high-temperature metallic phase is believed to be due to $\Delta < W$,[12] where $\Delta$ and W are the charge-transfer energy and the O *2p*-Ni *3d* hybridization strength or covalence bandwidth, respectively. As for the low-temperature insulating phase, the mechanism is still under debate though it was recently proposed that partial charge disproportionation among the Ni sites lowers the potential energy and gaps the Fermi surface.[13-16] Both theoretical and experimental studies reveal that the band gap opening due to charge disproportionation is at a few hundred millivolts,[4,14] leading to a moderate change in conductivity of one to two orders of magnitude.

One of the widely studied nickelates is $NdNiO_3$ (NNO),[7,11,17] which exhibits MIT at ~200 K in bulk. Earlier studies focused on band filling control by chemical doping and electric-field effect.[18-21] Hole-doping by partial substitution of $Ca^{2+}$ or $Sr^{2+}$ for $Nd^{3+}$ resulted in the closure of the charge-transfer gap and lowered $T_{MI}$.[19] Field-effect control of MIT in NNO films is challenging because of the high carrier concentration of ~ $10^{22}$ $cm^{-3}$ in the metallic phase at room temperature.[20] Recently, bandwidth-controlled MIT was realized in NNO films by varying the charge transfer energy and covalence bandwidth using strain.[7,22] Furthermore, dimensionality effect has also been investigated and the crossover from 3D to 2D can stabilize room temperature insulating phase.[23] However, these effects only exist in ultra-thin highly strained films, dramatically limiting their applications in various devices.

Oxygen vacancy, intrinsic to perovskite oxides, plays a major role in the electronic and magnetic properties of strongly correlated systems.[24-30] It effectively acts as double electron donors,[24] tuning the valance states of B-site transitional metal cations in $ABO_3$ perovskite oxides. Here, we demonstrate that an insulating phase can be stabilized at room temperature by controlling oxygen vacancy concentration in NNO, with a resistance modulation of greater than six orders of magnitude. Further study revealed that this phenomenon is universal for perovskite nickelates, and similar effect has been observed in $LaNiO_3$ (LNO) and $SmNiO_3$ (SNO) films. To demonstrate the potential application, heterojunctions of $NNO/SrTiO_3/Nb-SrTiO_3$ (Nb-STO) have been prepared and a photovoltaic power conversion efficiency (PCE) of about 0.61% has been observed. These results clearly demonstrate the importance of oxygen vacancy in nickelates and opens new pathways to look for alternative photovoltaic materials.

**EXPERIMENTAL SECTION**

**Sample preparation and Structure characterization.** Nickelates (LNO, NNO and SNO) films were deposited on (001)-oriented STO, and 0.5 wt% Nb-doped $SrTiO_3$ (Nb-STO) substrates by using pulsed laser deposition (PLD) method. During the deposition, the substrate temperature was kept at 630 $^{\circ}$C and growth oxygen pressure was kept at 40 Pa. After deposition, we raised the oxygen pressure to 10 kPa and then the samples were cooled down to room temperature. To introduce oxygen vacancies, the films were annealed in the PLD chamber under $5*10^{-4}$ Pa vacuum for 20 min at 120 $^{\circ}$C, 150 $^{\circ}$C, 180 $^{\circ}$C, 210 $^{\circ}$C , 240 $^{\circ}$C , 270 $^{\circ}$C and 300 $^{\circ}$C, respectively. To eliminate oxygen vacancies, the films were annealed in the PLD chamber under $10^4$ Pa oxygen environments for 20 min at 300 $^{\circ}$C. The surface morphology of the films was examined by atomic force microscope (AFM). X-ray diffraction measurements were performed using a Rigaku SmartLab instrument.

**Electrical and magnetic measurements.** In-plane transport property of the nickelates films were investigated using a 14 T PPMS (physical properties measurement system, Quantum Design) in the temperature range of 1.9-400 K. Linear four point geometry with Pt top electrodes was used. An Oxford Instruments system was also used to measure the sheet resistances of vacuum annealed nickelates films. In this case, a Keithley 220 current source and a Keithley 2700 multimeter were used. When the sheet resistance of the vacuum annealed nickelates film was above 200 MΩ, two point measurements were conducted using a pA metre/direct current (DC) voltage source (Hewlett Package 4140B) on a low noise probe station. The magnetic field was applied along the film surface when measuring the magnetic properties in the PPMS with a vibrating sample magnetometer.

**Hall measurements.** The 80 u.c. NNO films after annealing at different temperatures were patterned into hall bar structures using optical lithography and Ar-ion beam etching technique. The width and length of the bar is 50 μm and 550 μm, respectively. An external perpendicular magnetic field was applied while an AC current was used to measure the Hall resistivity. The AC frequency was 103 Hz and the duration was 5 s. All hall measurements were performed using PPMS system at room temperature.

**X-ray photoelectron spectroscopy (XPS) and ultraviolet photoelectron spectroscopy (UPS) measurements.** XPS and UPS measurements were performed in a home-built UHV multi-chamber system with base pressure better than $1 \times 10^{-9}$ torr. The XPS source was monochromatic Al Kα with photon energy at 1486.7 eV. The UPS source was from a helium discharge lamp (hν = 21.2 eV). The photoelectrons were measured using an electron analyzer (Omicron EA125). All the measured spectra were adjusted with Shirley type background, shifted with respect to carbon (C-C) XPS peak position due to charging.

**Absorption spectra and photovoltaic effect measurements.** The absorption spectra measurements were performed by using a UV-Vis-NIR spectrometer (200-2000 nm). In order to avoid the influence of STO absorption edge (~380 nm), we only varied the wavelength from 400 to 2000 nm. The scan rate was about 160 nm/min. The optical bandgap $E_g$ can be determined from the absorption coefficient α, calculated according to Tauc plot following $\alpha E = A(E-E_g)^n$, where α is the absorption coefficient, E is the photon energy, A is a constant, and n is equal to 1/2 or 2 for direct- or indirect-gap materials, respectively. In our case, the presence of clear linear relationships in the $(\alpha E)^2$ versus E curves indicate that the bandgap is direct. For the photovoltaic measurements, square Au top electrodes with a diameter of 400 $\mu$m×400 $\mu$m and a thickness of ~6 nm were deposited on the nickelates films through a metal shadow mask. The photovoltaic properties of the Au/NNO/Nb-STO and Au/NNO/STO/Nb-STO heterojunctions were investigated under the illumination of Xenon light with power density of 150 mW/cm$^2$.

**RESULTS AND DISCUSSION**

A set of NNO films ranging from 10 to 240 unit cells (u.c.) were grown on STO substrates using pulsed laser deposition. Although $T_{MI}$ varies with film thickness, all of the as-grown films show metallic behavior at room temperature (**Figure S1**).

**Oxygen vacancy induced metal-insulator transition and huge resistance switching.** **Figure 1**a shows the time dependence of sheet resistance ($R_{sheet}$) of a 40 u.c. NNO film measured at 27 ℃ and 127 ℃ in PPMS system (See details in **EXPERIMENTAL SECTION**). No obvious change is observed at 27 ℃, while it gradually increases with time at 127 ℃, possibly due to the creation of oxygen vacancies. For a better understanding, the samples were annealed at various temperatures between 120 ℃ to 300 ℃ for 20 min in vacuum and the resulting R-T curves are presented in **Figure 1**b. It is clear that, even for a low annealing temperature (120 ℃), the as-grown room temperature metallic behavior of the film has changed to insulating (**Figure S2**), suggesting that very small change of oxygen content can induce MIT in NNO films. Upon increasing the annealing temperature, the amount of oxygen vacancies is expected to increase, accompanied by increasing change of the $R_{sheet}$. After annealing in vacuum at 300 ℃ (V(a)300℃), more than six orders of magnitude change is observed in $R_{sheet}$ for the NNO films. Furthermore, annealing the sample in O$_2$ at 300 ℃ for 20 minutes (O(a)300℃) recovers the room temperature metallic behavior. This whole process can be cycled again (**Figure 1**c). The induced insulating phase is very stable at room temperature, and no change in the $R_{sheet}$ is observed after 10 hours (**Figure S3**).

Similar phenomenon has also been observed in LNO and SNO films (**Figure 1**d and **Figure S4**), indicating that it may exist in the whole RNiO$_3$ family. On the contrary, we have performed the same experiments on other conducting perovskite oxides, e.g., $La_{0.7}Sr_{0.3}MnO_3$ (LSMO) and $SrRuO_3$ (SRO). No obvious change in $R_{sheet}$ is observed as shown in **Figure 1**d.

**How do oxygen vacancies stabilize the insulating phase in $RNiO_3$?** For better understanding, below we focus on NNO to interpret the behind mechanism. Comparing the vacuum annealed films with the as-grown one, the (002) x-ray diffraction peak shifts towards smaller angle (**Figure 2**a), indicating the increase of $c$ lattice constant after vacuum annealing. Furthermore, the NNO films also change from black to semitransparent after vacuum annealing (the inset of **Figure 2**b). This is consistent with the band gap opening as revealed by the absorption spectra shown in **Figure 2**b. An optical gap of 1.86 eV was deduced for NNO films vacuum annealed at 300 ℃.

**Figure 2**c shows the hall resistivity ($\rho$) of the NNO films as functions of applied magnetic field at room temperature. The Hall coefficient ($R_H$) of as-grown film is $1.75\times10^{-4}$ $cm^3/C$, suggesting $p$-type conduction. Following $p^*=\frac{1}{eR_H}$ (where e is the elementary charge of $1.602\times10^{-19}$ C), we obtain an effective $p$-type carrier concentration ($p^*$) of $3.56\times10^{22}$ $cm^{-3}$ for as-grown films, which is consistent with previous reports.[20,31] Similar positive Hall coefficients have been reported for LNO and SNO films.[32,33] With increasing annealing temperature, the Hall coefficient remains positive and $p^*$ decreases as shown in **Figure 2**d. From **Figure 2**d, it can be seen that $c$ of the NNO film expands upon gradually increasing the vacuum annealed temperature, increasing from 3.7936 Å for the as-grown state to 3.8087 Å for the state of V(a)300℃. The modification of the lattice constants can be attributed to the influence of the loss of oxygen produced during the vacuum annealing process. Due to the loss of oxygen, Ni ions with the high valence state ($Ni^{3+}$) will vary to low valence state ($Ni^{2+}$) in order to satisfy the charge neutrality within each unit cell, thus leading to a larger Ni ionic radius. Regarding the antibonding nature of the $\sigma^*$ (O 2p-Ni 3d) bond,[13] Ni-O bonds expand and a larger lattice constant is observed. On the other hand, oxygen vacancies equal to act as a double electron donor, therefore, $p^*$ ($R_H$) gradually decreases (increases) upon increasing the vacuum annealed temperature. And about six orders of magnitude in $R_{sheet}$ modulation was observed. Similar to $R_{sheet}$, $c$ can be recovered to 3.7936 Å after annealing in $O_2$ environmental at 300 °C, indicating that the surface (the inset of **Figure 1**c) and the parent framework are not destroyed throughout the process.

In order to better understand the effect of oxygen vacancies, we turn to X-ray photoelectron spectroscopy (XPS) and ultraviolet photoelectron spectroscopy (UPS) measurements. From Ni 2p XPS (**Figure 3**a), the spectral line shapes of vacuum annealed samples indicate the gradual $Ni^{3+} \rightarrow Ni^{2+}$ valence change.[34] For example, the peak of Ni 2p3/2 at 853.9 eV reflects the dominance of $Ni^{2+}$ in NNO film vacuum annealed at 300 °C (the inset of **Figure 3**a).[35] Further analysis is presented in **Figure S5**. **Figure 3**b shows the valance band spectra weight of the NNO films obtained by UPS measurements. The valance band maximum (VBM) is determined by linear extrapolation of valance band onset.[36] For the vacuum annealed NNO films, VBM shifts from ~ 0.36 eV to ~ 0.95 eV below $E_F$. **Figure 3**c also reveals that the work function (WF) decreases with increasing vacuum annealing temperature, consistent with oxygen vacancies being donor dopants.

Combining the transport measurements and XPS/UPS results, the phase evolution of NNO films upon vacuum annealing is presented in **Figure 3**d. For the as-grown NNO films, the room temperature metallic behavior arises from the overlap of occupied O 2p and unoccupied Ni 3d bands.[37] Upon vacuum annealing, oxygen vacancies are introduced and the $Ni^{2+}/Ni^{3+}$ ratio gradually increases. Both valence and conduction bandwidths will be reduced due to the larger $Ni^{2+}$ ionic radius and Ni-O bond length ($d_{Ni-O}$).[38] The larger $d_{Ni-O}$ can also lead to the smaller Ni-O-Ni bond angle because the tolerance factor is reduced, further reducing the bandwidths. The narrowed bandwidths further lead to stronger Coulomb repulsion between the electrons (the intra-atomic d-d Coulomb repulsion energy U increases) and induce a large Mott–Hubbard splitting.[39] Both effects contribute to band gap opening and room temperature insulating phase is induced, as shown schematically in **Figure 3**d. This process is continuously tunable and reversible. Annealing the film in oxygen recovers the metallic behavior.

A direct consequence of the model we proposed is that the magnetization of NNO films should increase due to the high spin of $Ni^{2+}$ ($t_{2g}^6 e_g^2$),[14] as shown in **Figure 4**a. This was confirmed by measuring the magnetic property of an 80 u.c. NNO film before and after vacuum annealing at 300 °C using a vibrating sample magnetometer (VSM). Pronounced hysteretic loop is observed even at room temperature after vacuum annealing (**Figure 4**b). The saturation magnetization increases from 5.1 emu/cm$^3$ for the as-grown film to 51.3 emu/cm$^3$ after vacuum annealing at 300 ℃. The larger spin splitting is also a signal of larger U/W. Thus, introducing oxygen vacancies into NNO not only opens a gap at room temperature, but also increases the magnetization significantly, which provides new possibilities for nickelates in multifunctional devices.

**Applications of the nickelates-based heterojunction in photovoltaic cells.** Having clearly adjusted the bandgap by oxygen vacancies in the nickelates films suggests possible application in photovoltaic cells. We thus prepared heterojunctions made of *n*-type Nb-STO substrate and NNO films after different annealing treatments (see the structural schematic diagram shown in **Figure 5**a). **Figure 5**b shows the energy levels of NNO films with respect to that of Nb-STO. Semitransparent thin Au films are used as top electrodes. **Figure 5**c and **Figure 5**d display the J–V characteristics measured in dark and under 150 mW/cm$^2$ Xenon light illumination for Au/NNO(80 u.c.)/Nb-STO heterojunctions with NNO films after annealing at different temperatures. As shown in **Figure 5**c, the rectifying property of the heterojunction improves with increasing vacuum annealing temperature. The rectifying ratio defined as the ratio of forward-to-reverse currents at ±1.5 V, increases from 0.8 for the as-grown NNO to 200 for 300 ℃ annealed one. The improvement of the rectifying ratio can effectively reduce the dark saturation current ($I_0$) of the heterojunction, which should lead to larger open-circuit photovoltage ($V_{OC}$) according to the Shockley diode equation $V_{OC} = (nk_BT/q)\ln(I_L/I_0+1)$, where n is the ideality factor, $k_B$ is the Boltzmann constant, T is the temperature, q is the charge and $I_L$ is the light-generated current ($I_L$ is usually equal to $I_{SC}$, where $I_{SC}=J_{SC}$*the area of solar cell, $J_{SC}$ is the short-circuit photocurrent density).[40] On the other hand, higher annealing temperature leads to larger band gap and reduced $J_{SC}$ (**Figure 5**d). The very low $J_{SC}$ at the state of V(a)300℃ induced the drop off in $V_{OC}$. Optimized conditions can only be achieved by testing different combinations of film thickness, heterojunction and annealing condition. In **Figure 6**, we show the J-V curves of six devices under 150 mW/cm$^2$ Xenon light illumination. The detailed performances are summarized in Table 1. Comparing the *p-n* and *p-i-n* heterojunctions, where "*i*" refers to a 15 u.c. insulating STO layer, it is apparent that the insertion of a thin STO layer improves $V_{OC}$ without affecting $J_{SC}$,[41-43] resulting in the much improved PCE. Among them, the best performance is observed in Au/NNO(20 u.c.)/STO(15 u.c.)/Nb-STO [V(a)150℃] and the PCE is 0.61%. Note that these are preliminary results, and further optimization in materials selection and processing should improve the PCE even further.

## CONCLUSIONS

In summary, we have observed that oxygen vacancies open up band gaps in RNiO$_3$ films, likely due to the generation of Ni$^{2+}$ $t_{2g}^6 e_g^2$ states. Over six orders of magnitude change in the resistance is observed at room temperature together with the appearance of a large net magnetization. This discovery opens up new possibilities for perovskite nickelates. Potential application in photovoltaic cells has been demonstrated. A photovoltaic power conversion efficiency (PCE) of about 0.61% has been observed in the Au/NNO(20 u.c.)/STO(15 u.c.)/Nb-STO [V(a)150°C] heterostructure.

## ASSOCIATED CONTENT

### Supporting Information

This material is available free of charge online.

## AUTHOR INFORMATION

### Corresponding Authors

*Email: jlwang@ntu.edu.sg

### Notes

The authors declare no competing financial interest.

## ACKNOWLEDGMENTS

We thank Dr. Chen Shi and Prof. Tze Chien Sum (Division of Physics and Applied Physics, School of Physical and Mathematical Sciences, Nanyang Technological University, Singapore) for their help in the measurements of XPS and UPS. K. J. acknowledges support from the National Basic Research Program of China (No. 2014CB921001) and the National Natural Science Foundation of China (No. 11134012). We acknowledge financial support from Ministry of Education, Singapore under the Grant No. MOE2013-T2-1-052.

**FIGURES AND FIGURE CAPTIONS**

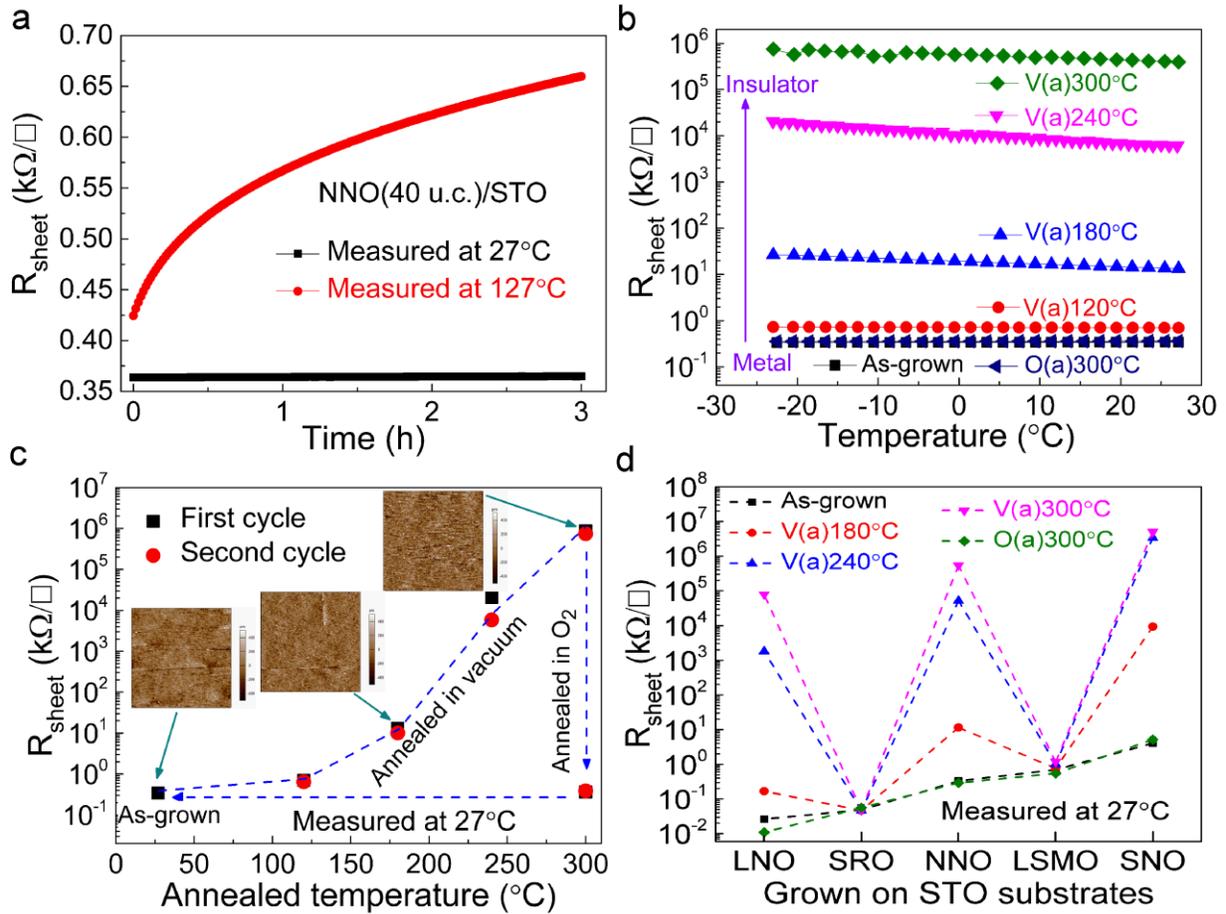

**Figure 1**. (**a**) Sheet resistance ($R_{sheet}$) of a 40 u.c. NNO film as a function of time measured at 27 ℃ and 127 ℃. (**b**) $R_{sheet}$ versus temperature for the 40 u.c. NNO film at various states, where V(a) and O(a) denote annealing in vacuum and annealing in O$_2$ environment, respectively. V(a)300℃ denotes annealing in vacuum at 300℃. (**c**) $R_{sheet}$ of the 40 u.c. NNO film as a function of annealing temperature. The black squares correspond to the first cycle, while the red circles correspond to the second cycle after the film was reoxidized by annealing in O$_2$ at 300℃. The inset shows the surface morphologies of NNO films at various states (left: the as-grown state; middle: V(a)180℃; right: V(a)300℃). The scan image size is $15\,\mu$m$\times15\,\mu$m. (**d**) Variation of room temperature $R_{sheet}$ of the 40 u.c. thick NNO film at various states is compared with other oxide films of the same thickness, LaNiO$_3$ (LNO), SrRuO$_3$ (SRO), La$_{0.7}$Sr$_{0.3}$MnO$_3$ (LSMO) and SmNiO$_3$ (SNO).

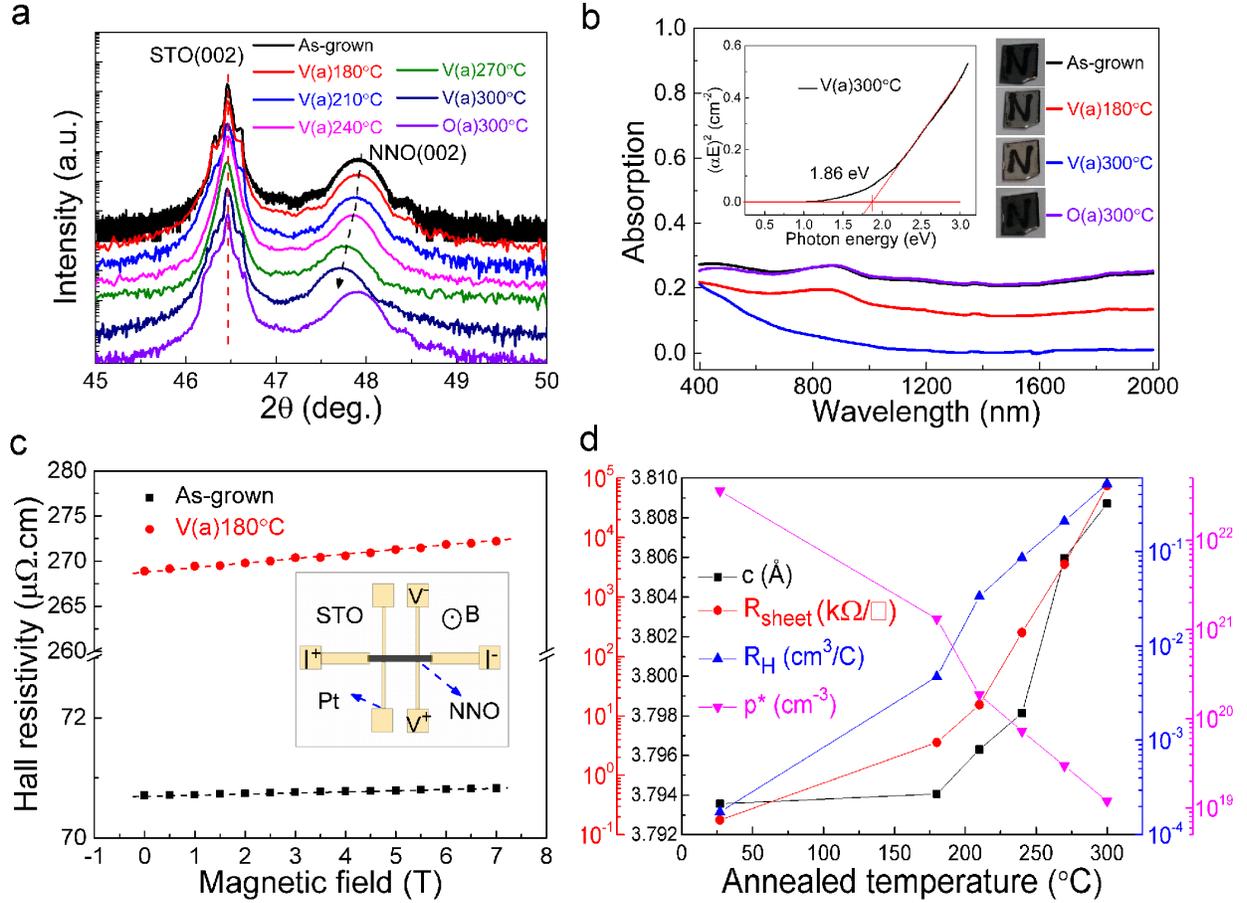

**Figure 2**. Room temperature XRD $\theta-2\theta$ scans around the (002) peaks (**a**) and optical absorption spectrum (**b**) for a 80 u.c. thick NNO films at various states. The insets in (**b**) reveal the optical band gap of NNO films after V(a)300℃ and the color change at various states. (**c**) Hall resistivity ($\rho$) as a function of magnetic field (B) for a 80 u.c. thick NNO films measured at room temperature. The inset is the schematic diagram of Hall effect measurement. (**d**) Room temperature $c$-axis lattice constant ($c$), $R_{sheet}$, hall coefficiency ($R_H$) and effective carrier concentration (p*) as functions of vacuum annealing temperature, respectively. $R_H$ is calculated using the formula $R_H = \frac{\rho(7T)-\rho(0T)}{70000} \times 10^8 cm^3/C$.

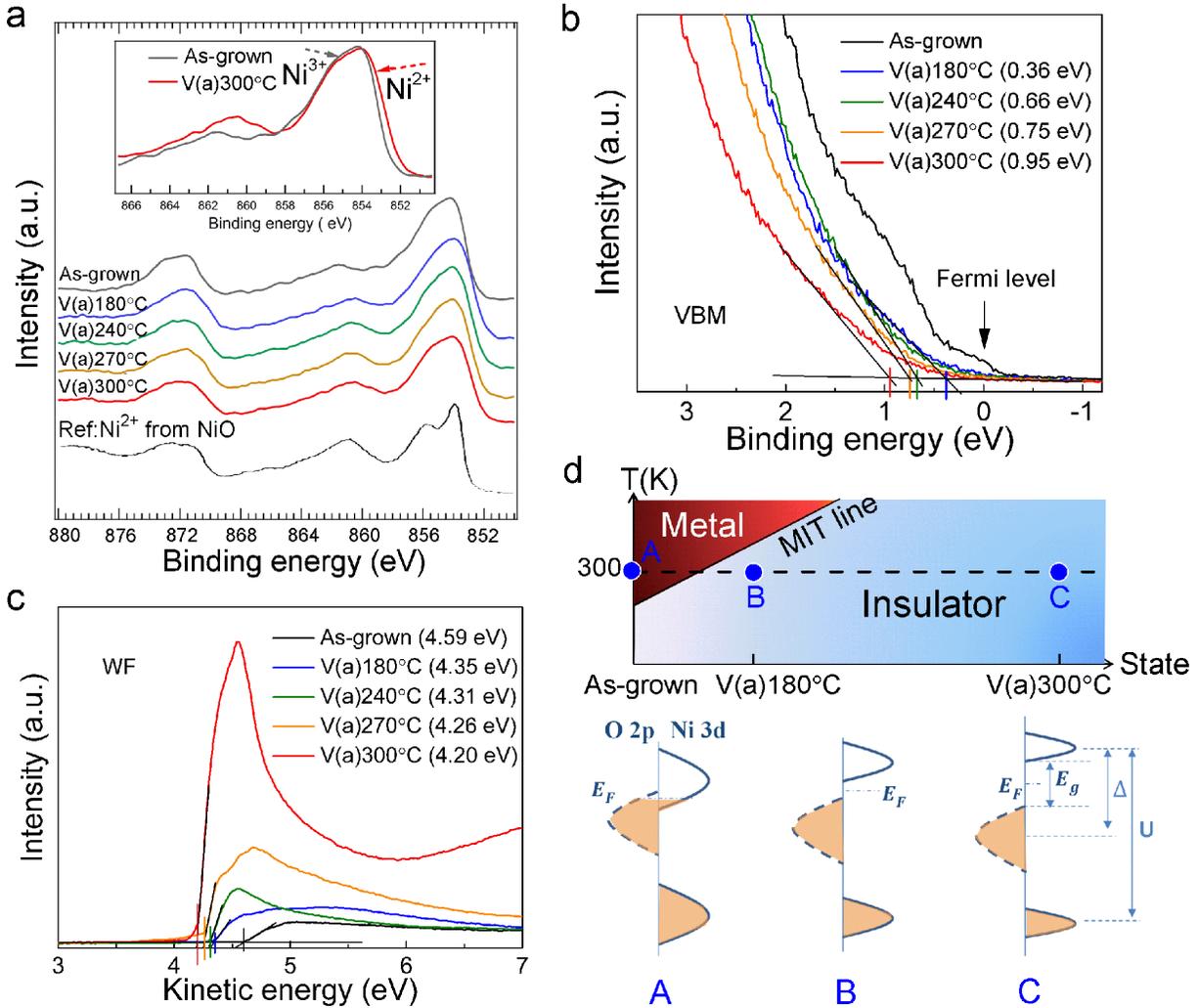

**Figure 3**. (**a**) Ni 2p XPS of as-grown and oxygen deficient 80 u.c. NNO films. The inset shows an enlargement for Ni 2p XPS of as-grown sample and that of V(a)300℃, which indicates an increase in $Ni^{2+}$ fraction after annealing in vacuum. Valence band maximum (VBM) (**b**) and work function (WF) (**c**) determined from the secondary electron onset in UPS measurements. (**d**) Schematic phase diagram of NNO film undergoing a room temperature metal-insulator transition due to oxygen vacancy. Schematic energy band diagrams for three different states are shown below the phase diagram. Here, Δ denotes the charge transfer energy, U is d-d Coulomb repulsion energy, $E_F$ is the Fermi level, and $E_g$ is the band gap.

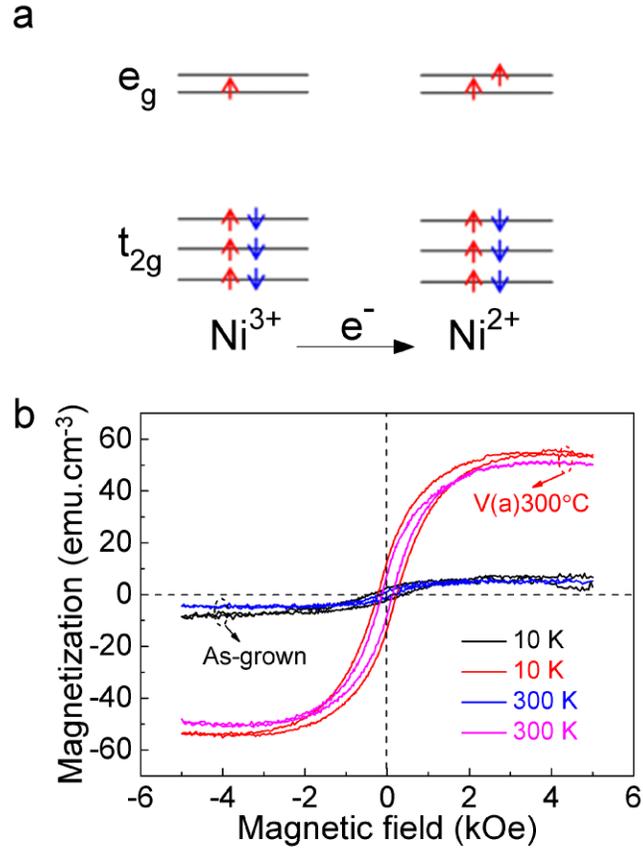

**Figure 4**. (**a**) The electron configuration of Ni 3d orbital in NNO is modifying by oxygen vacancy, with $Ni^{2+}$ ($t_{2g}^6 e_g^2$) replacing $Ni^{3+}$ ($t_{2g}^6 e_g^1$). (**b**) Magnetic hysteresis loops measured at 10 K and 300 K for the 80 u.c NNO film before and after vacuum annealing at 300 ℃.

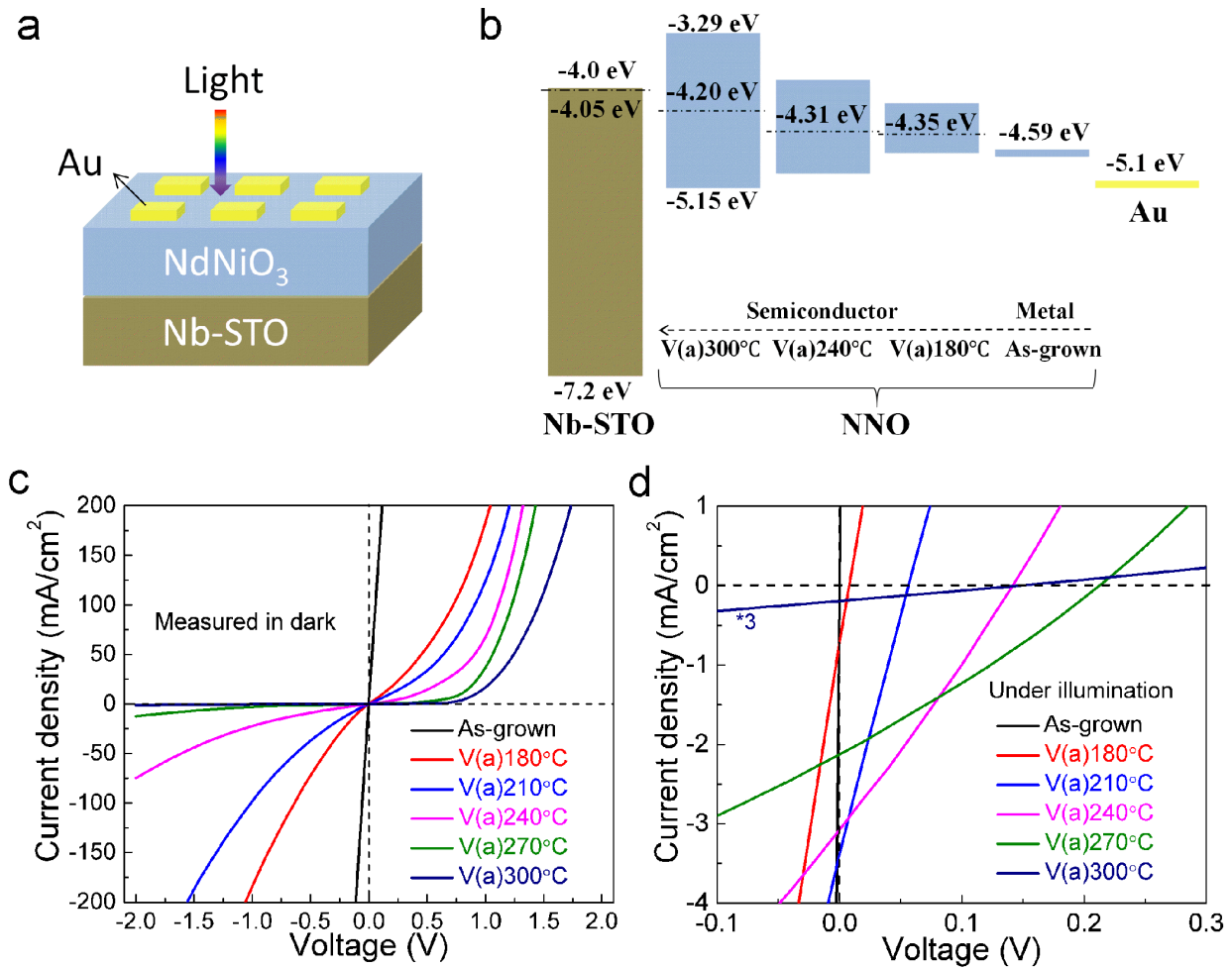

**Figure 5**. (**a**) Device architecture of the Au/NNO/Nb-STO heterojunctions. (**b**) Energy levels and work functions of the various components in the heterojunctions. The dashed lines denote the Fermi level. J–V characteristics of the Au/NNO(80 u.c.)/Nb-STO device measured in dark (**c**) and under 150 mW/cm$^2$ xenon light illumination (**d**) with NNO films after annealing at different temperatures.

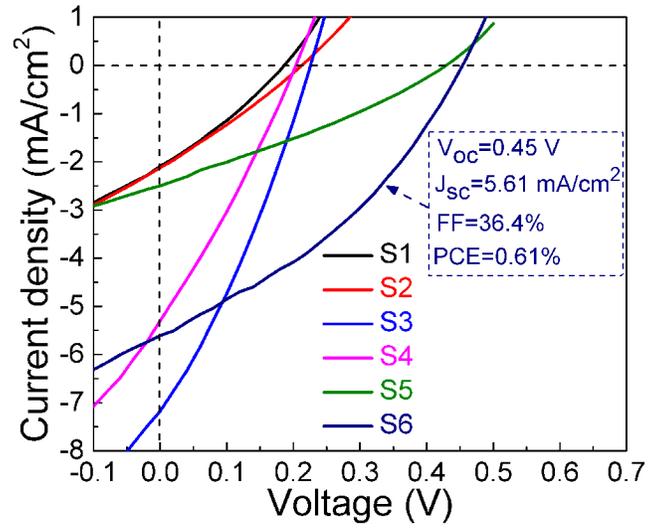

**Figure 6**. J–V characteristics of six different devices under 150 mW/cm$^2$ xenon light illumination. PCE is defined as PCE=$V_{OC}$*$J_{SC}$*FF/$P_{in}$, where FF is the filling factor, $P_{in}$ is the input power density (150 mW/cm$^2$). The details of the devices are shown in **Table 1**.

**TABLES AND TABLE CAPTIONS**

**Table 1.** Photovoltaic performance of six different devices.

| Name | Structure | $V_{OC}$ (V) | $J_{SC}$ (mA/cm$^2$) | FF (%) | PCE (%) |
|---|---|---|---|---|---|
| S1 | Au/NNO(120 u.c.)/Nb-STO [V(a)270°C] | 0.18 | 2.09 | 30.3 | 0.076 |
| S2 | Au/NNO(80 u.c.)/Nb-STO [V(a)270°C] | 0.21 | 2.12 | 26.6 | 0.079 |
| S3 | Au/NNO(40 u.c.)/Nb-STO [V(a)180°C] | 0.22 | 7.2 | 31 | 0.33 |
| S4 | Au/NNO(20 u.c.)/Nb-STO [V(a)150°C] | 0.2 | 5.32 | 26.6 | 0.19 |
| S5 | Au/NNO(80 u.c.)/STO/Nb-STO [V(a)270°C] | 0.43 | 2.51 | 29.8 | 0.21 |
| S6 | Au/NNO(20 u.c.)/STO/Nb-STO [V(a)150°C] | 0.45 | 5.61 | 36.4 | 0.61 |

# Supporting Information

Film thickness plays an important role in the properties of $RNiO_3$ films.[1-4] As shown in Figure S1a, all of the as-grown NNO films show metallic behavior at room temperature. Figure S1b shows the room temperature $R_{sheet}$ change after annealing as a function of NNO film thickness. The room temperature $R_{sheet}$ of 10 u.c. NNO film increases by six orders of magnitude after being annealed at 180 ℃ for 20 minutes. With further annealing at higher temperature, $R_{sheet}$ gradually saturates. For the thickest NNO film, although the room temperature metallic behavior is changed to insulating behavior after annealing in vacuum at 300 ℃, only two orders of magnitude modulation in $R_{sheet}$ at room temperature was observed. Moreover, after annealing in $O_2$ environment at 300 ℃, all of the NNO films recover the as-grown metallic state. It can be seen that the change of sheet resistance ($\Delta R_{sheet}$) between the as-grown state and after annealing in vacuum becomes larger with increasing vacuum annealing temperature. On the other hand, $\Delta R_{sheet}$ becomes smaller with increasing film thickness, indicating that oxygen vacancies are more likely created in thinner films.

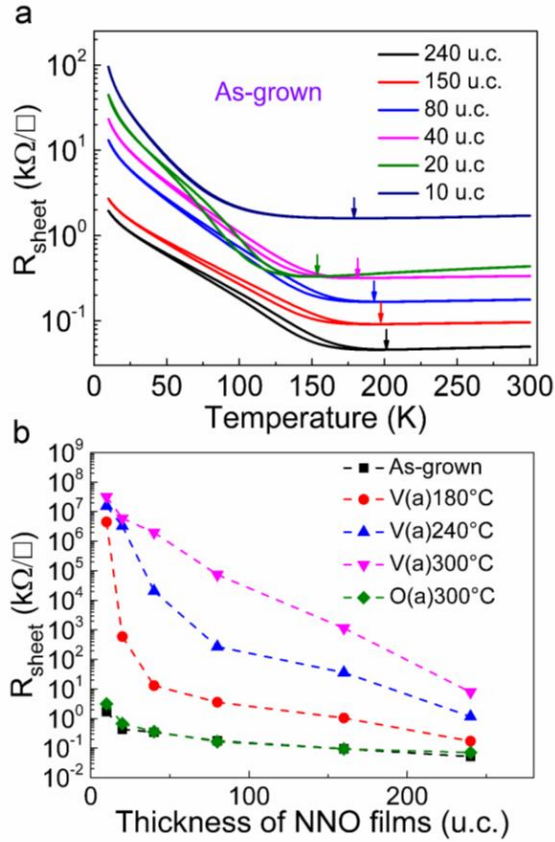

**Figure S1.** Thickness effect on the oxygen vacancy induced insulating phase. (**a**) $R_{sheet}$ versus temperature curves for NdNiO$_3$ (NNO) films of various thickness grown on STO. The arrows show the metal insulator transition temperature ($T_{MI}$) upon heating. (**b**) The effect of NNO film thickness on the room temperature $R_{sheet}$ change induced by oxygen vacancies.

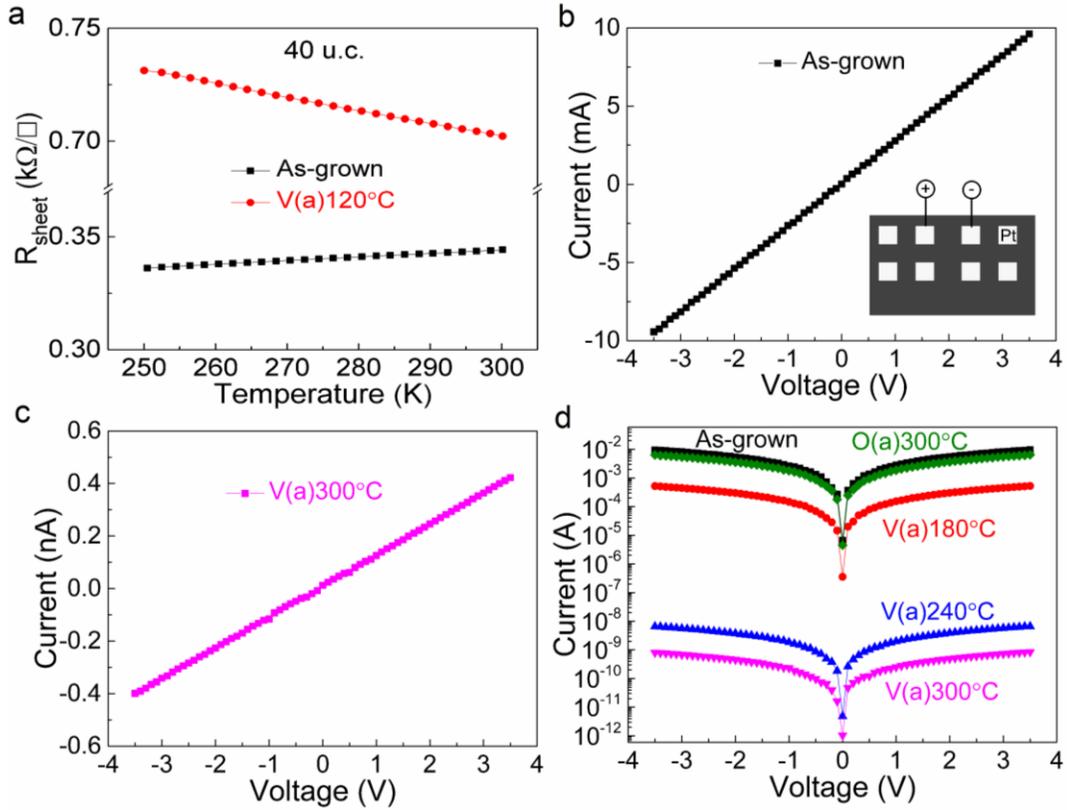

**Figure S2.** Ohmic contact. (**a**) $R_{sheet}$ versus temperature for 40 u.c. NNO films at the as-grown state and after vacuum annealing at 120 °C for 20 min. Clear metal-insulator transition is observed. (**b**) I-V curve of as-grown 40 u.c NNO film. The inset shows the schematic diagram for I-V measurements. (**c**) I-V curve of 40 u.c thick NNO film at the state of V(a)300°C. The good linear relationship indicates the Ohmic contact between Pt and NNO. (**d**) Logarithmic I-V characteristics for the NNO films at various states.

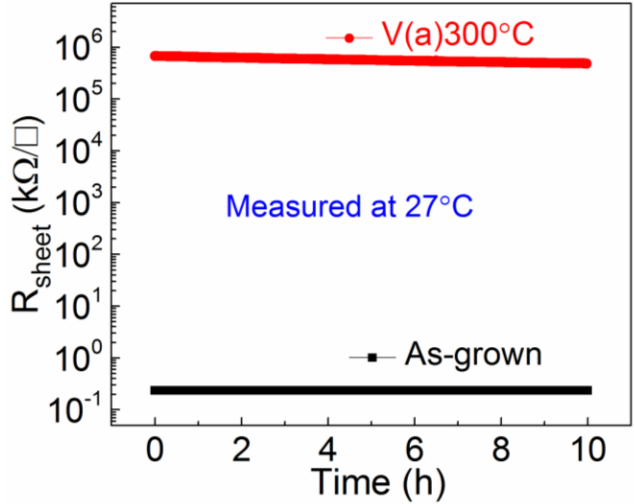

**Figure S3.** Retention characteristic of the NNO films at room temperature. The black squares correspond to the $R_{sheet}$ of as-grown film, while the red circles correspond to the $R_{sheet}$ of V(a)300°C film. No changes are observed in both cases after 10 hours.

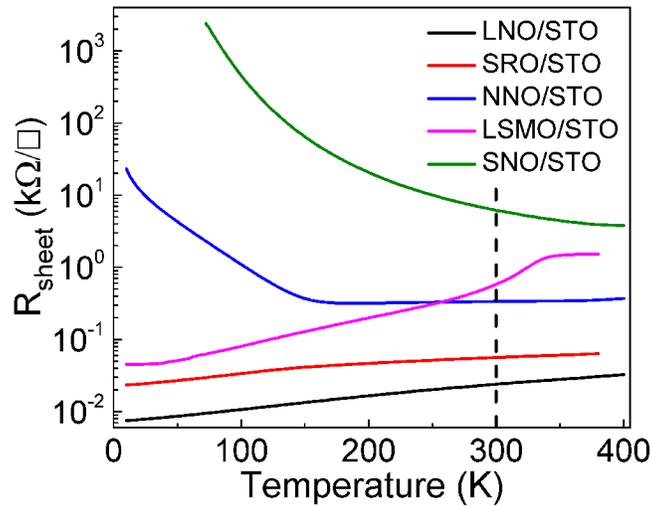

**Figure S4**. Comparison with other widely studied oxides. $R_{sheet}$ versus temperature curves for the as-grown $LaNiO_3$ (LNO), $SrRuO_3$ (SRO), NNO, $La_{0.7}Sr_{0.3}MnO_3$ (LSMO) and $SmNiO_3$ (SNO) films grown on STO substrate. The dash line shows the difference of their $R_{sheet}$ values at room temperature.

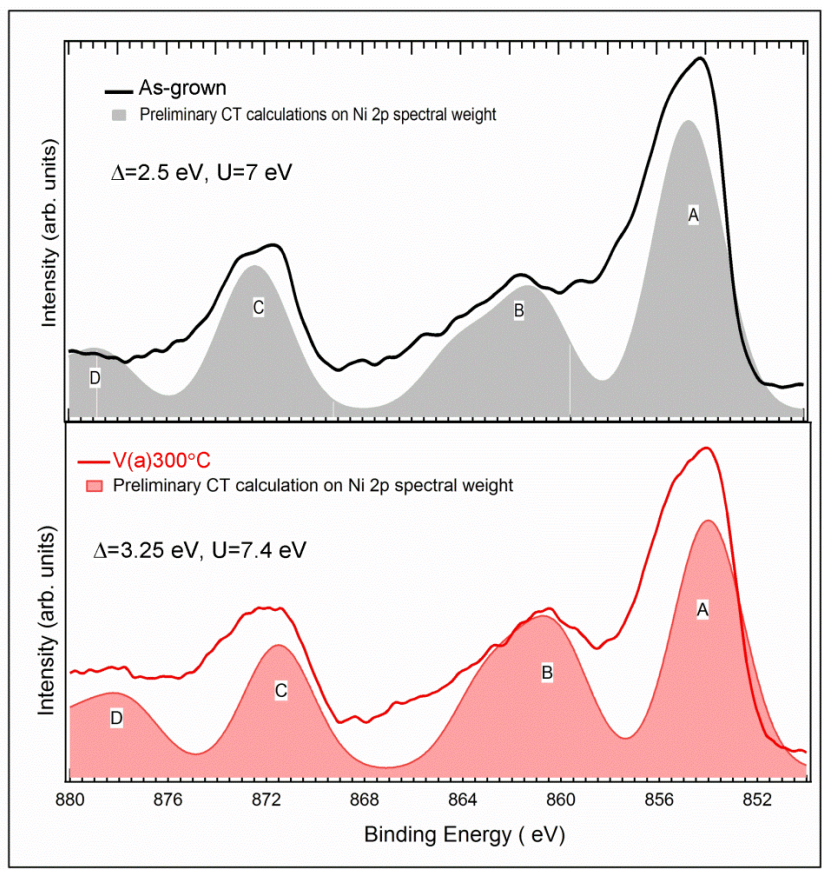

**Figure S5.** Preliminary charge transfer calculation based on XPS results. The Ni 2p XPS spectra containing peaks A and C due to spin orbit splitting and their satellite peaks B and D are shown in Figure S5. Satellites are often found in Ni based oxides and are ascribed to charge transfer (CT) effects from the ligand anions (O 2p) to the Ni 3d levels. These effects can be accounted for in the frame of a configuration interaction (CI) model, where the electronic states involved in the photoemission process are described by a linear combination of several configurations such as $3d^n$, $3d^{n+1}\underline{L}$, $3d^{n+2}\underline{L}^2$ ($\underline{L}$ represents a hole in the anion created by the CT to the *3d* cation orbitals).[5] The CT energy ($\Delta$) is defined by $=E(d^{n+1})-E(d^n)$. The positions of B and D satellites vary depending on $\Delta$ between the *p* and *d* orbitals involved in the CT process (here from O *2p* to Ni *3d*). U is the on-site coulomb repulsion between *d* electrons and $Q_{pd}$ is the intra-atomic coulomb interaction between *p* and *d* orbitals. The CT calculation has earlier been applied to Mn

based Ge system.[6] Similarly, CT calculation is applied to Ni *2p* spectra for as-grown and vacuum annealed samples, as shown in Figure S5. For the case of as-grown sample, the calculated CT (shaded area) curve (upper panel in Figure S5) is obtained with $\Delta = 2.5$ eV, $U = 7$ eV, and $Q_{pd} = 9$ eV. Calculated CT curve (shaded area) for the sample with the state of V(a)300°C has been obtained (lower panel in Figure S5) by setting $\Delta = 3.25$ eV, $U = 7.4$ eV, and $Q_{pd} = 10.6$ eV. These findings are consistent with those obtained on NiO systems.[7,8] $\Delta$, $U$, $Q_{pd}$ for the insulating state (vacuum annealed sample) is 0.75, 0.4 and 1.6 eV higher than the metallic state (as-grown sample) and it roughly scales the gap and hence discriminates the two systems. In addition to this, the spin orbit split of as-grown sample (17.7 eV) is quenched to 17.5 eV for the sample with the state of V(a)300°C, as a signature of transition to $Ni^{2+}$ of NiO-like system.